\documentstyle[12pt]{article}
\begin{document}
\begin{center} {\bf FORMS OF RELATIVISTIC DYNAMICS, CURRENT OPERATORS
AND DEEP INELASTIC LEPTON-NUCLEON SCATTERING}\end{center}
\vskip 1em
\begin{center} FELIX M. LEV\end{center}
\begin{center} {\it Laboratory of Nuclear Problems, Joint Institute
for Nuclear Research, Dubna, Moscow region 141980, Russia}\end{center}
\begin{center}(Contribution to the Conference on Perspectives
in Nuclear Physics at Intermediate Energies, 8-12 May 1995,
Trieste, Italy)\end{center}
\vskip 1em
\begin{abstract}
The three well-known forms of relativistic dynamics are unitarily
equivalent and the problem of constructing the current operators can
be solved in any form. However the notion of the impulse approximation
is reasonable only in the point form. In particular, the parton model
which is the consequence of the impulse approximation in the front form
is incompatible with Lorentz invariance, P invariance and T invariance.
The results for deep inelastic scattering based on the impulse
approximation in the point form give natural qualitative explanation
of the fact that the values given by the parton model sum rules
exceed the corresponding experimental quantities.
\end{abstract}

\flushleft{\bf 1. Forms of Relativistic Dynamics}\\
\vskip 0.5em

  The title of this section repeats the title of paper$^1$ in
which the notion of the forms of relativistic dynamics was introduced
for the  first time. In quantum field theory the four-momentum operators
${\hat P}^{\mu}$ ($\mu=0,...3$) and the representation generators of the
Lorentz group
${\hat M}^{\mu\nu}=-{\hat M}^{\nu\mu}$ ($\mu,\nu=0,...3$) are given
by integrals from the energy-momentum and angular momentum tensors
over a space-like (or light-like) hypersurface. Dirac$^1$ related
the different choices of the hypersurfaces to the different forms of
relativistic dynamics.

By definition, the description in the point form implies that the operators
${\hat M}^{\mu\nu}$ are the same as for noninteracting particles,
and thus interaction terms can be present only in the operators
${\hat P}$. If some operator is the same as for the noninteracting
system, we shall write this operator without "a hat". Therefore,
the point form is defined by the condition
${\hat M}^{\mu\nu}=M^{\mu\nu}$ and in the general case
${\hat P}^{\mu}\neq P^{\mu}$ for all $\mu$.

The description in the instant form implies that the
operators of ordinary momentum and angular momentum do not depend on
interactions, i.e. ${\hat {\bf P}}={\bf P}$, ${\hat {\bf M}}={\bf M}$
$({\hat {\bf M}}=({\hat M}^{23},{\hat M}^{31},{\hat M}^{12}))$ and
therefore interactions may be present only in ${\hat P}^0$ and the
generators of the Lorentz boosts ${\hat {\bf N}}=({\hat M}^{01},
{\hat M}^{02},{\hat M}^{03})$.

In the front form with the marked $z$
axis we introduce the + and - components of the 4-vectors as $p^+=
(p^0+p^z)/\sqrt{2}$, $p^-=(p^0-p^z)/\sqrt{2}$. Then we require that
the operators ${\hat P}^+,{\hat P}^j,{\hat M}^{12},{\hat M}^{+-},
{\hat M}^{+j}$ $(j=1,2)$ are the same as the corresponding free
operators and therefore interaction terms may be present only in the
operators ${\hat M}^{-j}$ and ${\hat P}^-$. We see that the front form
contains three interaction dependent generators while the point and
instant forms contain four such generators. However, in the front form
the operators ${\hat U}_P$ and ${\hat U}_T$ corresponding to space
reflection and time reversal necessarily depend on
interactions. This is clear from the relations
\begin{equation}
{\hat U}_PP^+{\hat U}_P^{-1}={\hat P}^-,\quad
{\hat U}_PM^{+j}{\hat U}_P^{-1}=-{\hat M}^{-j}\quad (j=1,2)
\label{1}
\end{equation}
and the analogous relations for the operator ${\hat U}_T$. At the same
time, in the point and instant forms we can choose
representations with ${\hat U}_P=U_P$ and ${\hat U}_T=U_T$.

 In paper$^1$ the instant form was related to the hypersurface
$t=0$, the front form to the hypersurface $x^+=0$, and the point
form to the hypersurface $t^2-{\bf x}^2<0,\,t>0$, but as argued
by Sokolov$^2$ the point form should be related to the hyperplane
orthogonal to the four-velocity of the system under consideration.
We shall not dwell on this
question but note that the choice of the hypersurface is not crucial.
This follows from the results of paper$^3$ in which the unitary
operators relating all the three forms were explicitly constructed.

Therefore {\it all the forms are unitarily equivalent and choosing a
form is only the matter of convenience but not the matter of principle}.

\flushleft{\bf 2. Constructing the Current Operators}\\
\vskip 0.5em

 Let ${\hat  J}^{\mu}(x)$ be the operator of the electromagnetic or
weak current where $x$ is the point in Minkowski space. The
translational invariance of the current operator implies that
\begin{equation}
{\hat  J}^{\mu}(x)=exp(\imath {\hat P}x){\hat  J}^{\mu}(0)
exp(-\imath {\hat P}x)
\label{2}
\end{equation}
This relation makes it possible to reduce the problem of seeking
${\hat  J}^{\mu}(x)$ to the problem of seeking ${\hat  J}^{\mu}(0)$. The
latter should satisfy Lorentz invariance which implies that
\begin{equation}
[{\hat M}^{\mu\nu},{\hat  J}^{\mu}(0)]=-\imath (g^{\mu\rho}
{\hat  J}^{\nu}(0)- g^{\nu\rho}{\hat  J}^{\mu}(0))
\label{3}
\end{equation}
where $g^{\mu\nu}$ is the metric tensor in Minkowski space.

 If the current operator satisfies the continuity equation
$\partial {\hat J}^{\mu}(x)/\partial x^{\mu}=0$ then, as follows from
Eq. (2), this equation can be written in the form
\begin{equation}
  [{\hat J}^{\mu}(0),{\hat P}_{\mu}]=0
\label{4}
\end{equation}

 Having constructed the current operators in some form and using the
unitary operators relating these forms we can construct the current
operators in any form. Therefore {\it the forms of dynamics are
equivalent not only in problems of calculating the binding energies,
scattering amplitudes etc. but also in problems of describing
electromagnetic or weak properties of strongly interacting systems}.

\flushleft{\bf 3. Electromagnetic and Weak Processes with Large Momentum
Transfer}\\
\vskip 0.5em

  Our intuition tells that when the momentum transfer in the
electromagnetic or weak process is very large then the virtual photon
or W-boson interacts only with one constituent, and the process is so
quick that the interaction between this constituent and the remnants
of the target can be neglected. People usually believe that the
mathematical expression of such a condition is
${\hat J}^{\mu}(x)=J^{\mu}(x)$ i.e. the current operator is the same
as for noninteracting particles. For calculations it suffice to use
this condition at $x=0$ and we assume that, by definition, the
impulse approximation (IA) is defined by the relation
\begin{equation}
{\hat J}^{\mu}(0)=J^{\mu}(0)
\label{5}
\end{equation}

  In which form Eq. (5) may be reasonable? It is important to note
that {\it if ${\hat J}^{\mu}(0)$ is free in some form then
${\hat J}^{\mu}(0)$ is not free in the other forms since the
unitary operators relating the forms are interaction dependent}.

 By looking at Eq. (3) we conclude that in the instant and front
forms none of the components of ${\hat J}^{\mu}(0)$ can be free since
some of the Lorentz group generators are interaction dependent.
Moreover, the choice (5) in the front form breaks also P and T
invariance, since, as noted in Sec. 1, the operators ${\hat U}_P$
and ${\hat U}_T$ are interaction dependent in this form. At the same time,
the choice (5) in the point form preserves Lorentz invariance, P
invariance and T invariance since the corresponding operators are
free in this form.

 Unfortunately, the problem is not so simple if ${\hat J}^{\mu}(0)$
should also satisfy the continuity equation (4). Let the initial state
$|i\rangle $ be the eigenstate of the four-velocity operator
with the eigenvalue $G'$ and the eigenstate of the mass operator with
the eigenvalue $M'$. Analogously, let the final state $|f\rangle$
be the eigenstate of the four-velocity and mass operators with the
eigenvalues $G"$ and $M"$ respectively. It is always possible to
consider the process in the reference frame in which
${\bf G}"+{\bf G}'=0$ (see paper$^4$ for details). We choose
the coordinate axes in such a way that $G_x'=G_y'=0$. Then, as
follows from Eq. (4), the $x$ and $y$ components of the operator
${\hat J}^{\mu}(0)$ are not constrained by the continuity equation
and the 0 and $z$ components satisfy the relation
\begin{equation}
G^{'0}(M"-M')\langle f|{\hat J}^0(0)|i\rangle=
G^{"z}(M"+M')\langle f|{\hat J}^0(0)|i\rangle
\label{6}
\end{equation}
In deep inelastic lepton-nucleon scattering $M'$ is the nucleon mass, and
in the Bjorken limit (when $G^{'0}=G^{"z}$ and $M" \gg M'$) the interaction
of the struck quark
with the remnants of the target can be neglected$^5$. Then it
follows from Eq. (6) that the condition (5) for all the components of the
operator ${\hat J}^{\mu}(0)$ is compatible with the continuity equation.

\flushleft{\bf 4. Brief Overview of the Parton Model Sum Rules for Deep
Inelastic Lepton-Nucleon Scattering}
\vskip 0.5em

 The present theory of deep inelastic lepton-nucleon scattering
is based on the operator product expansion (OPE)$^6$,
Altarelli-Parisi equations$^7$ and collinear expansion$^5$.
Many textbooks consider deep inelastic processes in the framework of
the Feynman parton model$^8$ proposed for explaining the phenomenon
of Bjorken Scaling$^9$. However there exist only a very few
cases when the results$^{5-7}$ agree with the
parton model. This occurs when the anomalous dimensions of the
Wilson coefficients are equal to zero and the momentum transfer $q$ is
so large that all higher twist effects (which are of order
$(M^{'2}/|q^2|)^n$) and the perturbative QCD corrections (of order
$\alpha_s(q^2)^n$, where $\alpha_s(q^2)$ is the QCD running coupling
constant and $n=1,2...$) can be neglected. We shall always assume that
$q$ indeed satisfies such properties.

 There exist three sum rules which agree with the parton model
at large $|q^2|$. These are the sum rules for the unpolarized
neutrino-nucleon deep inelastic scattering derived by Adler$^{10}$,
Bjorken$^{11}$, and Gross and Llewellyn Smith$^{12}$.
The existing experimental data do not make it possible to verify the
first and the second sum rules with good accuracy, and the precise
data recently obtained by the CCFR collaboration$^{13}$ show
that the experimental value of the Gross-Llewellyn-Smith sum $S_{GLS}$
is smaller than the value $S_{GLS}=6$ predicted by the parton model.
The analysis of the CCFR data in papers$^{14,15}$ shows that
actually $S_{GLS}=4.90\pm 0.16\pm 0.16$ and taking
into account the corrections of order $\alpha_s(q^2)$ and
$\alpha_s(q^2)^2$ is still insufficient to explain the above deficiency.

 A very nontrivial sum rule for the polarized electron-nucleon scattering
was derived by Bjorken$^{16}$. The left-hand-side of this sum
rule can be written in terms of the parton model, but the right-hand-side
is given not by the normalization integral for the nucleon wave function
(as in the above sum rules), but by the matrix element of the axial
charge operator. Therefore, the right-hand-side is determined
by low-energy physics and cannot be written in terms of the parton
model. One of the major problems in comparing the Bjorken sum rule with
the data is the problem of extracting the first moment of the neutron
polarized structure function $g_1(x)$ from the deuteron data. For this
purpose it is necessary to construct the deuteron electromagnetic
current operator satisfying Eqs. (2-4). Let us also note that some data
on the Gross-Llewellyn Smith and Bjorken sum rules were actually
obtained at not very large $|q^2|$ (a few $GeV^2$).

 Let us now consider the sum rules which one way or another are based on
the parton model.

 According to the recent precise results of the NMC group$^{17}$
the experimental value of the integral defining the Gottfried sum
rule$^{18}$ is equal to $0.235\pm 0.026$ instead of 1/3 in the
parton model.

 The EMC result$^{19}$ for the first moment of the proton
polarized structure function $g_1(x)$ is $\Gamma_p=0.126\pm 0.010
\pm 0.015$ while the Ellis-Jaffe sum rule$^{20}$ predicts
$\Gamma_p^{EJ}=0.171\pm 0.004$. The fact that the Ellis-Jaffe sum
rule gives the value considerably exceeding the corresponding
experimental quantity was also confirmed at this conference$^{21,22}$.

 The most impressive results of the parton model sum rules are those
concerning the quark contribution to the nucleon's momentum and spin.
The first sum rule (see, for example, the discussions in the
textbooks$^{23}$ says that quarks carry only a half of the nucleon
momentum, and this fact is usually considered as one of those which
demonstrates the existence of gluons. The second sum rule known as
"the spin crisis" says that the quark contribution to the nucleon
spin is comparable with zero (a detailed discussion of the spin crisis
can be found, for example, in papers$^{24-26}$), and the
SMC and E143 groups$^{21,22}$ estimate this contribution as 25\%.
Of course, these results are not in direct contradiction with
constituent quark models since the latter are successful only at low
energies. Nevertheless, our experience can be hardly reconciled with
the fact that the role of gluons is so high.

 The above discussion gives all grounds to conclude that the values
given by the parton model sum rules systematically exceed the
corresponding experimental quantities.

\flushleft{\bf 5. Qualitative Explanation of the Deficiency in the Parton Model
Sum Rules}\\
\vskip 0.5em

 As shown by several authors (see, for example, papers$^{27}$ and
references cited therein), the parton model is a consequence of the
IA (i.e. Eq. (5)) in the front form of dynamics.
Meanwhile, as noted in Sec. 3, the current operator satisfying Eq. (5)
in the front form does not properly commute with the Lorentz boost
generators and the operators ${\hat U}_P$ and ${\hat U}_T$. Therefore
{\it the parton model is incompatible with Lorentz invariance, P
invariance and T invariance}. The question arises what is the
quantitative extent of the violation of these symmetries in the parton
model? To answer this question we have to compare the results obtained
by using the current operator satisfying the above symmetries with
the results of the parton model.

 If we assume that at large $|q^2|$ the IA is valid, then, as noted in
Sec. 3, the only choice is to use the IA in the point form. Let us also
note that if the operator ${\hat J}^{\mu}(0)$ satisfies Eq. (5), it
has transitions only between single-particle states and there are
no Z-diagrams. In addition, the notion of the infinite momentum frame
is not necessary in the point form, since the Lorentz boost generators
are free in this form and the components of the Fock column are not
mixed by the Lorentz boost transformations.

 A detailed comparison of the results in the IA for the front and
point forms have been carried out in paper$^{28}$. Below we
qualitatively explain what is the difference between the corresponding
calculations.

  Let us consider a system of $N$ particles with the masses $m_i$
($i=1,...N$). It is not important whether the number $N$ is finite
or infinite. Let ${\bf k}_i$ be the momenta of these particles in their
c.m. frame such that ${\bf k}_1+...{\bf k}_N=0$. The energy of particle
$i$ in the c.m. frame is equal to
$\omega_i({\bf k}_i)=(m_i^2+{\bf k}_i^2)^{1/2}$, and the {\it free} mass
of the N-body system is equal to $M_0=\omega_1({\bf k}_1)+...+
\omega_N({\bf k}_N)$.

 Instead of ${\bf k}_i$ we can introduce the variables
$({\bf k}_{i\bot},\xi_i)$, where ${\bf k}_{i\bot}$ is the projection of
${\bf k}_i$ onto the plane $xy$ and
\begin{equation}
\xi_i=\frac{\omega_i({\bf k}_i)+k_i^z}{M_0({\bf k}_1,...{\bf k}_N)}
\in (0,1)
\label{7}
\end{equation}
For simplicity we shall suppress the spin variables in the internal
wave function $\chi({\bf k}_{1\bot},\xi_1,...{\bf k}_{N\bot},\xi_N)$.
Then we can choose this function in such a way that the normalization
condition is
\begin{eqnarray}
&&\int\nolimits |\chi({\bf k}_{1\bot},\xi_1,...{\bf k}_{N\bot},\xi_N)|^2
\delta^{(2)}({\bf k}_{1\bot}+...+{\bf k}_{N\bot})\cdot\nonumber\\
&&\delta(\xi_1+...+\xi_N-1)\prod_{i=1}^{i=N}d^2{\bf k}_{i\bot}d\xi_i =1
\label{8}
\end{eqnarray}

 Consider the process in the reference frame where the momentum $P'$ of
the initial nucleon is such that ${\bf P}'_{\bot}={\bf q}_{\bot}=0$,
and $P^{'z}$ is positive and very large. Then the calculations in the
front form$^{27,28}$ give the well-known result that if the
virtual photon is absorbed by quark $i$ then $\xi_i=x$ where
$x=|q^2|/2(P'q)$ is the Bjorken variable. At the same time, the
calculations in the point form$^{28}$ show that the relation
between $\xi_i$ and $x$ is
\begin{equation}
M_0({\bf k}_{1\bot},\xi_1,...{\bf k}_{N\bot},\xi_N)(1-\xi_i)=M'(1-x)
\label{9}
\end{equation}

 This relation shows that the difference between the parton model and
our approach depends on the difference between the free mass $M_0$ and
the real nucleon mass $M'$. In the nonrelativistic approximation
$M_0=M'$ since the binding and kinetic energies can be neglected and
both quantities are equal to $m_1+...+m_N$. However there is no reason
to believe that the nucleon is the nonrelativistic system.

 Equation (9) makes it possible to explicitly determine the $\xi_i$ as
a function of the other internal variables and $x$:
$\xi_i=\xi_i({\bf k}_{\i\bot},int_i,x)$ where $int_i$ means the internal
variables for the system $1,...i-1,i+1,...N$. The details are given in
paper$^{28}$. It is important to note that when $x\in [0,1]$,
$\xi\in [\xi_i^{min},1]$ where $\xi_i^{min}$ is the function
$\xi_i({\bf k}_{\i\bot},int_i,x=0)$. {\it The important conclusion is
that the data on deep inelastic lepton-nucleon scattering do not make
it possible to determine the $\xi_i$ distribution of quarks at
$0<\xi_i<\xi_i^{min}$. Of course, the quantity $\xi_i^{min}$ can be
determined only if some concrete model of the nucleon is assumed}.

 Now the fact that the parton model sum rules give values exceeding the
corresponding experimental quantities can be qualitatively explained as
follows. In the parton model the integration over $x\in [0,1]$ can be
replaced by the integration over $\xi_i\in [0,1]$, and the right-hand-side
of the parton model sum rules becomes proportional to the normalization
integral (8). At the same time, in our approach the integration over
$x\in [0,1]$ can be transformed to the integration over $\xi_i\in
[\xi_i^{min},1]$. Therefore the right-hand-side of the corresponding sum
rules does not contain the contribution of $\xi_i\in [0,\xi_i^{min}]$,
and it is natural to expect that the corresponding integral is smaller
than the normalization integral (8). The details are given in paper$^{28}$.

\flushleft{\bf 6. Discussion}\\
\vskip 0.5em

 We have proposed a simple qualitative explanation of the deficiency
in the parton model sum rules. Our considerations show that the parton
language is not convenient since the $x$ distribution is not uniquely
related to the quark distribution over $\xi_i$.

 The parton model gives a clear explanation of the phenomenon of
Bjorken Scaling. Let us note however that Bjorken Scaling is in fact
only a consequence of the fact that the dimensionless structure functions
$F_i(x,M^{'2}/|q^2|)$ and $g_i(x,M^{'2}/|q^2|)$ are not singular when
$|q^2|\rightarrow \infty$. This can take place in different models;
in particular this takes place in model$^{28}$.

 Purists can say that only those sum rules are of significance which
can be derived from the OPE, i.e. the sum rules$^{10-12,16}$
(originally these sum rules were derived assuming that the algebra of
the equal time commutation relations for the current operators is the
same as for the free current operators, but this is a special case of
the OPE). The OPE is used for the perturbative calculation of the
product of the currents entering into the expression for the hadronic
tensor:
\begin{equation}
W^{\mu\nu}=\frac{1}{4\pi}\int\nolimits e^{\imath qx} \langle N|
{\hat J}^{\mu}(x){\hat J}^{\nu}(0)|N\rangle d^4x
\label{10}
\end{equation}
where $|N\rangle$ is the state of the initial nucleon. Let us note
however that the OPE has been proved only in the framework of the
perturbation theory which does not apply to the bound state problem.
As noted above, the current operator ${\hat J}^{\mu}(x)$ should
satisfy the relations (2-4). At the same time the nucleon state
$|N\rangle$ is the eigenstate of the four-momentum operator ${\hat P}$
with the eigenvalue $P'$ and the eigenstate of the spin operator which
is composed from the operators ${\hat M}^{\mu\nu}$. Therefore the relation
between ${\hat J}^{\mu}(x)$ and $|N\rangle$ is highly nontrivial, and it
is not clear in advance whether the Wilson coefficients for the
product of the currents entering into Eq. (10) can be sought by using the
expansion in $\alpha_s(q^2)$ while the state $|N\rangle$ entering into the
same equation is not affected by such an expansion.

 Let us also note that for the unambiguous verification of the
sum rules$^{10-12,16}$ they should be extracted directly
from the experimental data at large $|q^2|$, not using the $|q^2|$
evolution determined from the OPE or Altarelli-Parisi
equations. The above considerations show that such experimental data
are crucial to solve the problem whether perturbative QCD applies to
deep inelastic lepton-nucleon scattering. Our considerations also
pose the problem whether the role of gluons in the
deep inelastic lepton-nucleon scattering observables is so high as
usually believed. Of course, this problem deserves investigation.

\vskip 1em
\flushleft{\bf 7. Acknowledgments}\\
\vskip 0.5em
\begin{sloppypar}
 The author is grateful to A.Buchmann, F.Coester, A.E.Dorokhov,
S.B.Gerasimov,
S.Gevorkyan, I.L.Grach, A.V.Efremov, B.L.Ioffe, A.B.Kaidalov, V.A.Karmanov,
N.I.Kochelev,
E.A.Kuraev, L.A.Kondratyuk, S.Kulagin, E.Leader, G.I.Lykasov, L.P.Kaptari,
A.Makhlin, S.V.Mikhailov, I.M.Narodetskii, N.N.Nikolaev,
V.A.Novikov, E.Pace, M.G.Sapozhnikov, G.Salme, S.Simula, N.B.Skachkov,
O.V.Teryaev, Y.N.Uzikov  and  H.J.Weber for valuable discussions
and to S.J.Brodsky, S.D.Glazek, L.L.Frankfurt and M.P.Locher for useful
remarks. This work was supported by grant No. 93-02-3754 from the Russian
Foundation for Fundamental Research.
\end{sloppypar}
\vskip 1em
\flushleft{\bf 8. References}\\
\vskip 0.5em
\flushleft{1. P.A.M.Dirac, {\it Rev. Mod. Phys.} {\bf 21} (1949) 392}.\\
2. S.N.Sokolov, in {\it Problems in High Energy Physics
and Quantum Field Theory} (IHEP, Protvino, 1984) p. 85.\\
3. S.N.Sokolov and A.N.Shatny, {\it Teor. Mat. Fiz.}
{\bf 37} (1978) 291.\\
4. F.M.Lev, {\it Annals of Physics} {\bf 237} (1995) 355.\\
5. R.K.Ellis, W.Furmanski and P.Petronzio, {\it Nucl. Phys.}
{\bf B212} (1983) 29.\\
6. K.G.Wilson, {\it Phys. Rev.} {\bf 179} (1969) 1499;
K.G.Wilson and W.Zimmerman, {\it Commun. Math. Phys.} {\bf 24} (1972) 87.\\
7. G.Altarelli and G.Parisi, {\it Nucl. Phys.} {\bf B126}
(1977) 298.\\
8. R.P.Feynman, {\it Phys. Rev. Lett.} {\bf 23} (1969) 1415;
J.D.Bjorken and E.A.Paschos, {\it Phys. Rev.} {\bf 185} (1969) 1975.\\
9. J.D.Bjorken, {\it Phys. Rev.} {\bf 179} (1969) 1547.\\
10. S.L.Adler, {\it Phys. Rev.} {\bf 142} (1966) 1144.\\
11. J.D.Bjorken, {\it Phys. Rev.} {\bf 163} (1967) 1767.\\
12. D.J.Gross and C.H.Llewellyn Smith, {\it Nucl. Phys.}
{\bf B14} (1969) 337.\\
13. CCFR Collab. P.Z.Quintas et. al. {\it Phys. Rev. Lett.}
{\bf 71} (1993) 1307.\\
14. A.L.Kataev and A.V.Sidorov, Report CERN-TH 7235/94
(CERN, Geneva, 1994); Report E2-94-344 (JINR, Dubna, 1994).\\
15. A.E.Dorokhov, {\it Pisma ZHETF} {\bf 60} (1994) 80.\\
16. J.D.Bjorken, {\it Phys. Rev.} {\bf 147} (1966) 1467.\\
17. New Muon Collaboration. M.Arneodo et al. {\it Phys. Rev.}
{\bf D50} (1994) R1.\\
18. K.Gottfried, {\it Phys. Rev. Lett.} {\bf 18} (1967) 1174.\\
19. EM Collaboration, J.Ashman et al., {\it Nucl. Phys.}
{\bf 328} (1989) 1.\\
20. J.Ellis and R.L.Jaffe, {\it Phys. Rev.} {\bf D9} (1974) 1444;
{\bf D10} (1974) 1669.\\
21. G.Baum, The report of the SM Collaboration to this conference.\\
22. K.Griffioen, The report of the E143 Collaboration to this
conference.\\
23. F.E.Close, {\it An Introduction to Quarks and Partons}
(Academic Press, London-New York, 1979); F.J.Yndurain, {\it Quantum
Chromodynamics. An Introduction To The Theory of Quarks and Gluons}
(Springer Verlag, New York - Berlin - Heidelberg - Tokyo, 1983).\\
24. J.Ellis and M.Karliner, Report CERN-TH-6898/93
(CERN, Geneva, 1993).\\
25. B.L.Ioffe, Report ITEP 61-94 (ITEP, Moscow, 1994).\\
26. A.E.Dorokhov and N.I.Kochelev, {\it Paricles and Nuclei}
{\bf 26} (1995) 5.\\
27. H.J.Weber, {\it Phys. Rev.} {\bf D49} (1994) 3160;
Z.Dziembowski, C.J.Martoff and P.Zyla, {\it Phys. Rev.} {\bf D50} (1994)
5613.\\
28. F.M.Lev, hep-ph 9501348.\\
\end{document}